\documentclass[runningheads,anonymous]{llncs}

\usepackage[T1]{fontenc}
\usepackage{subcaption}
\usepackage{graphicx}
\usepackage{amsmath}

\usepackage{amsthm}
\usepackage{amssymb}
\usepackage{url}

\begin{document}
\title{Steering Large Text-to-Image Model for Abstract Art Synthesis: Preference-based Prompt Optimization and Visualization}
\titlerunning{Steering Large Model for Deterministic Abstract Art Synthesis}
%
\author{
Aven-Le Zhou \and
Wei Wu \and
Yu-Ao Wang \and
Kang Zhang}
\authorrunning{A.L. Zhou et al.}
\institute{
The Hong Kong University of Science and Technology (Guangzhou), P.R. China\\
\email{aven.le.zhou@gmail.com}}
\maketitle              

\begin{abstract}

With the advancement of neural generative capabilities, the art community has increasingly embraced GenAI (Generative Artificial Intelligence), particularly large text-to-image models, for producing aesthetically compelling results. However, the process often lacks determinism and requires a tedious trial-and-error process as users often struggle to devise effective prompts to achieve their desired outcomes. This paper introduces a prompting-free generative approach that applies a genetic algorithm and real-time iterative human feedback to optimize prompt generation, enabling the creation of user-preferred abstract art through a customized ``Artist Model.'' The proposed two-part approach begins with constructing an Artist Model capable of deterministically generating abstract art in specific styles, e.g., Kandinsky's Bauhaus style. The second phase integrates real-time user feedback to optimize the prompt generation and obtains an ``Optimized Prompting Model,'' which adapts to user preferences and generates prompts automatically. When combined with the Artist Model, this approach allows users to create abstract art tailored to their personal preferences and artistic style.

\end{abstract}
\section{Introduction}

Generative Artificial Intelligence (GenAI) has emerged as a powerful tool for creating diverse forms of content, including text, sounds, images, and videos, demonstrating its transformative potential to enhance and extend human creativity \cite{beyond_prompt,dalle,Rombach_Blattmann_Lorenz_Esser_Ommer_2022,text2image_diffusion,yu2022scaling}. One of the most promising applications are large text-to-image models \cite{best_prompt}. Solutions such as DALL-E 2 \cite{dalle} and Stable Diffusion \cite{Rombach_Blattmann_Lorenz_Esser_Ommer_2022} employ joint image-text embedding techniques, e.g., CLIP \cite{clip} and diffusion models \cite{diffusion_model}, to generate realistic and aesthetically pleasing contents. These models have become popular for creating digital images \cite{VQGAN} and artworks \cite{creativity_of_text2image,Rombach_Blattmann_Ommer_2022}.

The text-to-image art (TTI-art) community has embraced various tools, from open-source models like Stable Diffusion to commercial platforms like Midjourney. This diverse community includes both amateurs and professionals, spanning novice creators and seasoned artists. As these models grow in capability and accessibility, digital art generated through text-to-image technology progresses rapidly and is on the brink of becoming a mainstream artistic medium \cite{creativity_of_text2image}.

\subsection{Prompting in Text-to-Image Generation}
Large models demonstrate unparalleled capabilities in synthesizing high-quality and diverse images and text \cite{dalle}. However, these models operate in a universal and general manner, lacking customization to match the user-specific preferences within a given reference set \cite{dreambooth}. Even with detailed descriptions, the text embedding space cannot fully capture the desired preferences, often producing only variations that may not align with user expectations \cite{clip}.

Common in text-to-image generation systems is the ability for users to create digital images and artworks by writing prompts in natural language. This process, known as prompt engineering \cite{prompt_engineering}, prompt programming \cite{prompt_programming}, prompt design \cite{prompt_design}, or simply prompting \cite{creativity_of_text2image}, enables users to generate content without requiring deep technical knowledge of the underlying technologies. However, the interaction between the user and AI often prioritizes the model's requirements over the human experience  \cite{multi_concept}. When creating art from the text, users inevitably relinquish some degree of control to the AI \cite{Galanter_2016}.

This restricted human-AI interaction often disregards users’ intuition and needs, making prompt generation in these systems feel arbitrary \cite{multi_concept,best_prompt}. While large text-to-image models can produce aesthetically pleasing results, their reliance on unstructured, open-ended text inputs frequently forces users into a trial-and-error approach \cite{prompt_engineering}. To refine their image generation process, users must continuously iterate on new prompts—a method described as ``random and unprincipled''. Ultimately, this chaotic and non-deterministic process makes prompting an ongoing challenge for end users.

\subsection{Controlling Non-determinism}

Chaos, in the context of digital art—particularly in generative art such as randomness and emergence—is not always perceived as negative \cite{art_emergence_random,emergence_genart}. Instead, it often presents opportunities that artists actively embrace \cite{ten_questions}. Generative artists employ processes with a degree of autonomy to create all or part of an artwork \cite{Galanter_2016}. Computer-programmed art, for example, intentionally incorporates chaos, such as randomness, as a central element of its creation process \cite{Bailey_2018}. This chaos or non-determinism often leads to emergence, producing unexpected and interesting results that exceed initial expectations \cite{emergence_genart}. Exploring non-deterministic artworks in a controlled manner and identifying the ``sweet spot'' between full control and total chaos is not new. Kovach describes generative art as a continual search for the ideal balance between complete control and total chaos \cite{Kovach_2018}.

When dealing with non-deterministic text-to-image models, it is essential to consider how generative art experience can inform the development of such AI technology and the evolving role of artists in this process. We argue that understanding and leveraging the history of generative art offers valuable insights into how one can better collaborate with GenAI. To support the controlled application of non-determinism in GenAI from an artistic perspective, we examine how ``traditional'' generative art approaches, such as procedural modeling and genetic algorithm (GA), can be adapted to large text-to-image models for painterly content creation.

\subsection{Steering Large Text-to-Image Model for Abstract Art Synthesis}

Within the TTI-art community, Low-Rank Adaptation of Large Language Models (LoRA) \cite{Hu_lora} has inspired numerous derivative applications to customize large text-to-image models. Given the popular and powerful nature of GenAI as a new artistic medium, insights from the art community can provide significant value. Building on these experiences, we adapt several techniques from the TTI-art community to customize a large text-to-image model with a specific artist style, referred to as the Artist Model. We term our approach ``semantic injection,'' a method for encoding a describable artist’s style into the large model.

\begin{figure}[h]
    \centering
    \includegraphics[width=\textwidth]{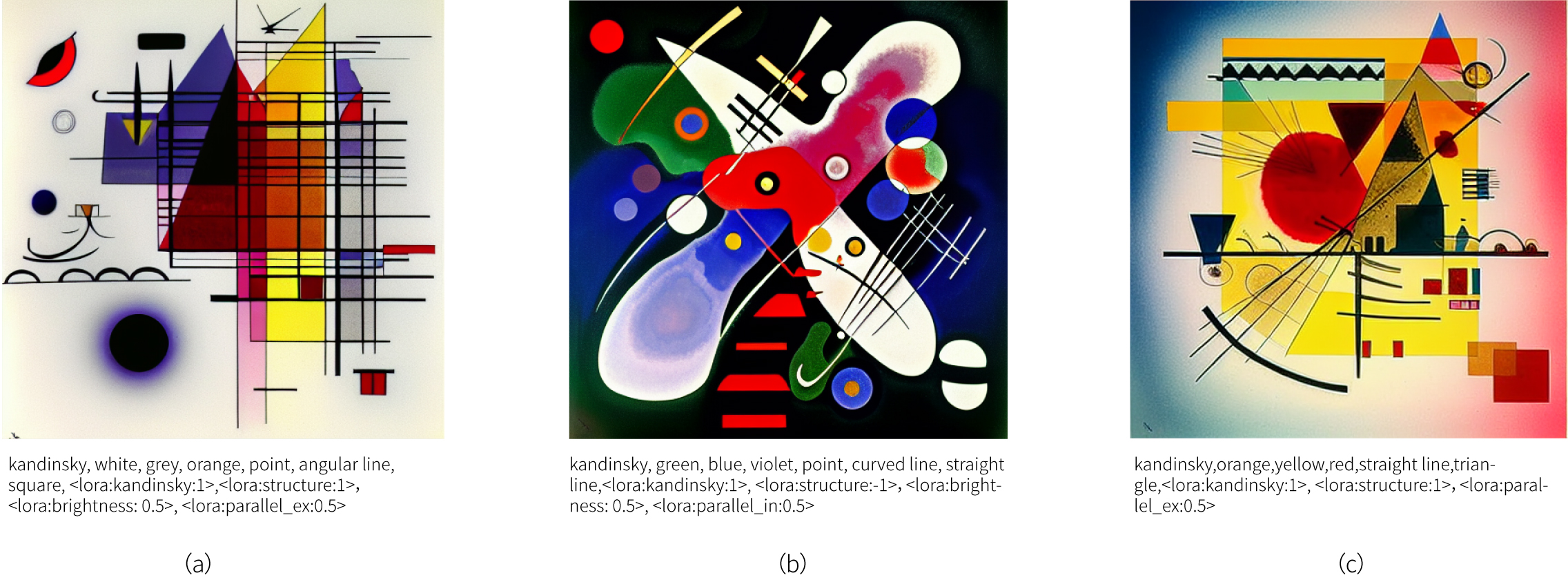}
    \caption{Artist Model (of Kandinsky Bauhaus Style) through Semantic Injection.}
    \label{fig:prompting}
    \end{figure}

\begin{figure}[h]
    \centering
    \includegraphics[width=\textwidth]{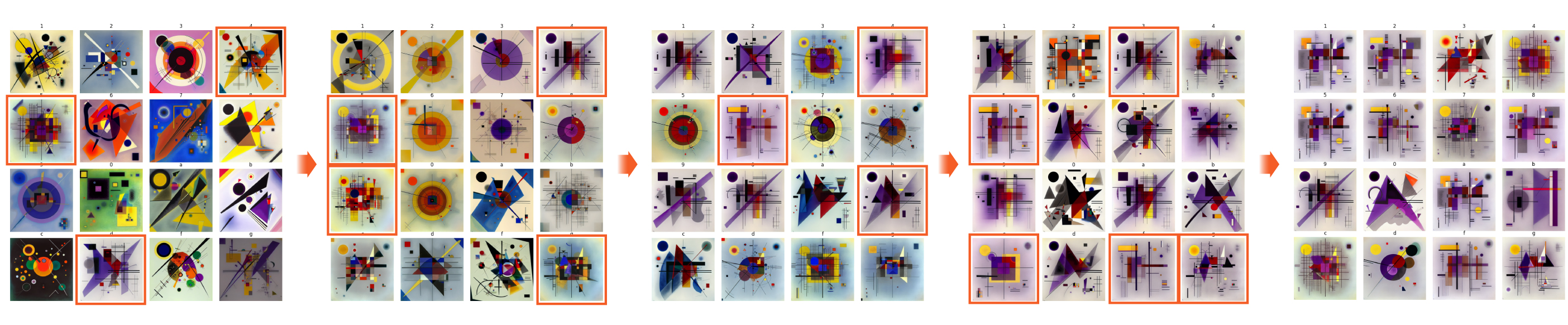}
    \caption{Genetic Prompting Optimization with a User's Real-time Feedback Leads to the ``Optimized Prompting Model'' in Five Iterations.}
    \label{fig:demo}
    \end{figure}

To further integrate insights from generative art, such as evolutionary generation, we propose applying a Genetic Algorithm combined with real-time iterative human feedback to optimize prompt generation. Fig \ref{fig:demo} shows this optimization process results in an ``Optimized Prompting Model,'' which eliminates the need for manual prompt engineering but automatically prompts with the user's preferences.

By combining the Optimized Prompting Model with the Artist Model, we offer a solution for abstract art synthesis using large text-to-image models incorporating user preferences. The Optimized Prompting Model operates autonomously, requiring no additional user input, and generates abstract art tailored to the user's preferences—completely eliminating the need for manual prompting. Our contributions can be summarized as follows:

\begin{enumerate}
    

    \item We propose a two-part approach that allows users to automatically create abstract art based on their preferences, eliminating the need for explicit prompting. This approach combines ``semantic injection,'' which customizes the large text-to-image model, with ``genetic prompting optimization,'' which employs real-time iterative feedback to generate prompts automatically.
        

    \item We construct a text-to-image dataset in the Kandinsky Bauhaus style and implement the proposed approach as an interactive, open-sourced system.
    
\end{enumerate}

\begin{figure}[h]
    \centering
    \includegraphics[width=\textwidth]{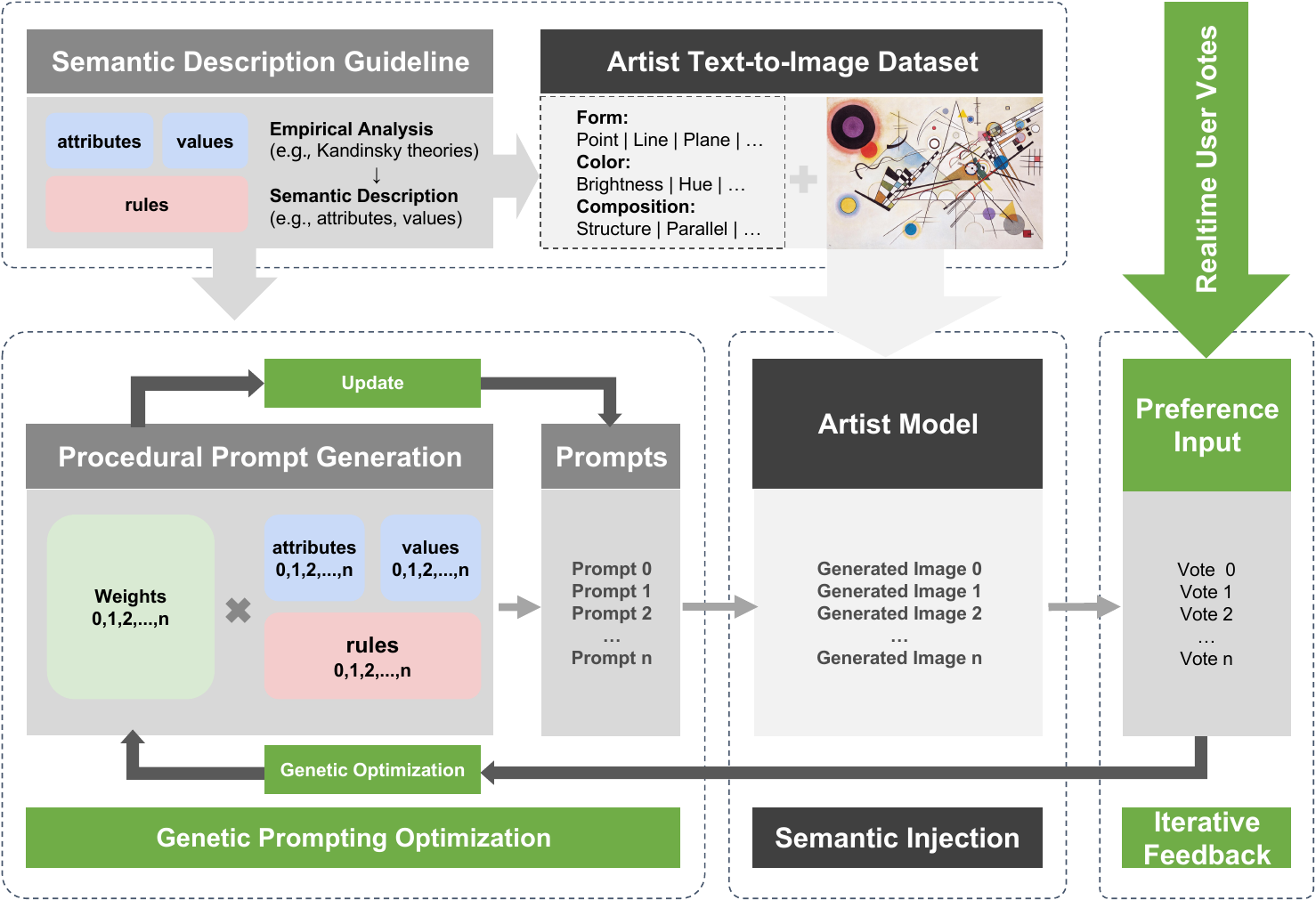}
    \caption{Genetic Prompting Optimization with Real-time Iterative Feedback based on the ``Artist Model'' obtained from \textit{Semantic Injection}.}
    \label{fig:genetic_optimization}
    \end{figure}

This paper presents the proposed approach, its methods, implementation experiments, and discussions. Sec \ref{relatedwork} positions the proposed approach and its methods within the current research landscape while Sec \ref{methods} explains our two primary methods: semantic injection for the large text-to-image customization and genetic prompting optimization. Sec \ref{dataset_kandinsky}, \ref{artistmodel_exp}, and \ref{promptingmodel_exp} provide a detailed account of the implementation process, including the experiments involved in constructing the dataset, developing the Artist Model, and creating the Optimized Prompting Model, as well as building the interactive system that delivers prompting-free user experiences for generating abstract art tailored to user preferences.

Fig \ref{fig:genetic_optimization} illustrates the process of our approach and experiments. We establish a semantic descriptive guideline in Sec \ref{dataset_kandinsky} and construct a text-to-image dataset in the Kandinsky Bauhaus style. Sec \ref{artistmodel_exp} applies the semantic injection method to customize the Artist Model with the dataset, forming the foundation for deterministic image synthesis. Sec \ref{promptingmodel_exp} implements Genetic Prompting Optimization through procedural prompt generation based on the semantic descriptive guideline and applies genetic optimization on the prompt generation through iterative user feedback. Ultimately, in Sec \ref{experience}, we design the interactive system and interface. In Sec \ref{discussion} and Sec \ref{limitation}, we reflect on the study's findings, discuss its broader implications, and address limitations and potential directions for future research.





\section{Related Work}\label{relatedwork}
\subsection{Text-to-Image Models Personalization}

Numerous studies have explored techniques for customizing large text-to-image models to specific objects or styles. For instance, DreamBooth personalizes text-to-image diffusion models by fine-tuning them with a small set of images \cite{dreambooth}, enabling the synthesis of photorealistic images while preserving the key features of the subject. Similarly, \cite{Wei_Zuo_2023} introduces a learning-based encoder that facilitates fast and accurate customized text-to-image generation. The LoRA method addresses the challenge of parameter efficiency by reducing the number of trainable parameters for downstream tasks and freezing the pre-trained model weights during adaptation \cite{Hu_lora}. Meanwhile, \cite{Feng_He_Fu_2023} proposes a training-free method for guiding diffusion models, enhancing attribute-binding and compositional capabilities. Additionally, \cite{Wolf_2023} presents a training-free strategy, FABRIC, which utilizes the self-attention layer in diffusion models to optimize content customization.

\subsection{Human Preference and Feedback}
Several studies have explored how to model human preferences and incorporate human feedback into text-to-image generation. For example, \cite{Wu_Sun_Zhu_Zhao_Li_2023} addresses the challenge of aligning text-to-image models with human preferences by employing a human preference classifier, which adapts the model to generate more preferred images. Similarly, \cite{Tang_Rybin_Chang_2023} improves the quality of images produced by a diffusion generative model using human ranking feedback. Meanwhile, \cite{Hao_Chi_Dong_Wei_2023} proposes a prompt adaptation framework that enhances text-to-image models by incorporating user input. However, this framework adapts to the model-preferred prompt rather than generating user-preferred prompts. Finally, \cite{best_prompt} presents a human-in-the-loop approach that employs a genetic algorithm to identify the optimal combination of prompt keywords. While similar to our method, this approach focuses exclusively on discovering the best prompt for large-scale models.

\section{Method} \label{methods}


\subsection{Semantic Injection}    
Building on open-source research efforts like LoRA, the TTI-art community has developed numerous effective variations, including FastLoRA \cite{Cloneofsimo}  and DiffLoRA \cite{DiffLoRA}. Our work adapts and extends these techniques to propose the concept of ``semantic injection,'' a method for customizing large text-to-image models.

\subsubsection{FastLoRA}
The FastLoRA technique optimizes the balance between quality and computational efficiency when fine-tuning stable diffusion models. Built upon research efforts such as Pivotal Tuning, Dreambooth, and Textual Inversion, the Fast LoRA technique adjusts the Transformer model's attention layers by adding low-rank matrices to the model's weights to achieve significant changes in model behavior by modifying a small number of parameters. For the pre-trained weight matrix $W \in \mathbb{R}^{n\times m}$, the adjusted weight matrix can be represented as $W' = W + \Delta W = W+AB^{T}$, where $A\in \mathbb{R}^{n\times d}$ and $B\in \mathbb{R}^{m\times d}$, $n$ being the dimension of the original weight matrix and $d$ being the low-rank factor, usually much smaller than $d$. The residual $\Delta W$ is thus decomposed into smaller-size matrices $A$ and $B$, allowing efficient model tuning instead of updating larger $W$. Furthermore, this technique only fine-tunes parts of the transformer model, specifically the self-attention heads, to reduce the final model size. 

\subsubsection{DiffLoRA}
The DiffLoRA technique is a multi-step operation that utilizes LoRA and two images: the original image A and the manipulated image B. Initially, we fine-tune the original stable diffusion model with image A using LoRA until it over-fits, resulting in Lora A. We then combine this over-fitting Lora A with the original model to create model A. Subsequently, we repeat the same process using model A (not the original model) and image B, generating the over-fitting LoRA B and model B. Finally, the differences between model B and the original model are captured to define the DiffLoRA, effectively encoding critical differences between the two inputs as continuous attribute values.

By combining FastLoRA and DiffLoRA, semantic injection efficiently encodes both discrete and continuous attribute values into large models. This process integrates various fine-tuned LoRA models with the original model to create a customized model capable of deterministic text-to-image generation, referred to as the Artist Model.

\subsection{Genetic Prompting Optimization and Visualization}

The genetic algorithm (GA), an evolutionary algorithm, mimics the natural evolutionary process through selection, crossover, and mutation operations on individuals \cite{Holland_1992}. With the Artist Model, we frame prompt generation as an optimization problem and apply GA with real-time iterative feedback for evolutionary optimization. Specifically, we define the generative function as $G: C \to I$, where $C$ represents chromosomes indicating prompts, and $I$ denotes individuals in the population corresponding to the generated images from the Artist Model. 

In this approach, individuals represent text-to-image generations, while chromosomes correspond to user prompts composed of genes encoding features of artistic style. We assume that users maintain consistent tendencies in their aesthetic preferences throughout the process and use them to evaluate the fitness of the individuals accordingly. Following Darwin's theory of ``Survival of the fittest'' \cite{Grefenstette_1993}, prompts that generate higher-fitness images are selected as survivors, leading to an Optimized Prompting Model that aligns with user preferences. 

To illustrate the dynamic evolution of prompts, we integrate multiple visualization techniques. The generated prompts and corresponding abstract art images are displayed side-by-side, offering users a comprehensive view of their preferences across both text and image domains. At each iteration, radar charts, bar charts, and stream graphs are used to effectively visualize the iterative changes in prompts.

\section{Dataset: Kandinsky Bauhaus Style}\label{dataset_kandinsky}

Kandinsky's artistic style reached a prominent phase characterized by geometric elements upon his return to Germany and his subsequent tenure at the Bauhaus in 1922. This period (1922-1933) also marked the publication of his significant work, \textit{Point and Line to Plane}. In this work and others, Kandinsky thoroughly explained his theories on color, form, and composition. Alongside his remarkable corpus of paintings executed under the guidance of these theories, Kandinsky Bauhaus style presents a perfect yet challenging subject for implementing our proposed approach.

\begin{figure}[h]
    \centering
    \includegraphics[width=\textwidth]{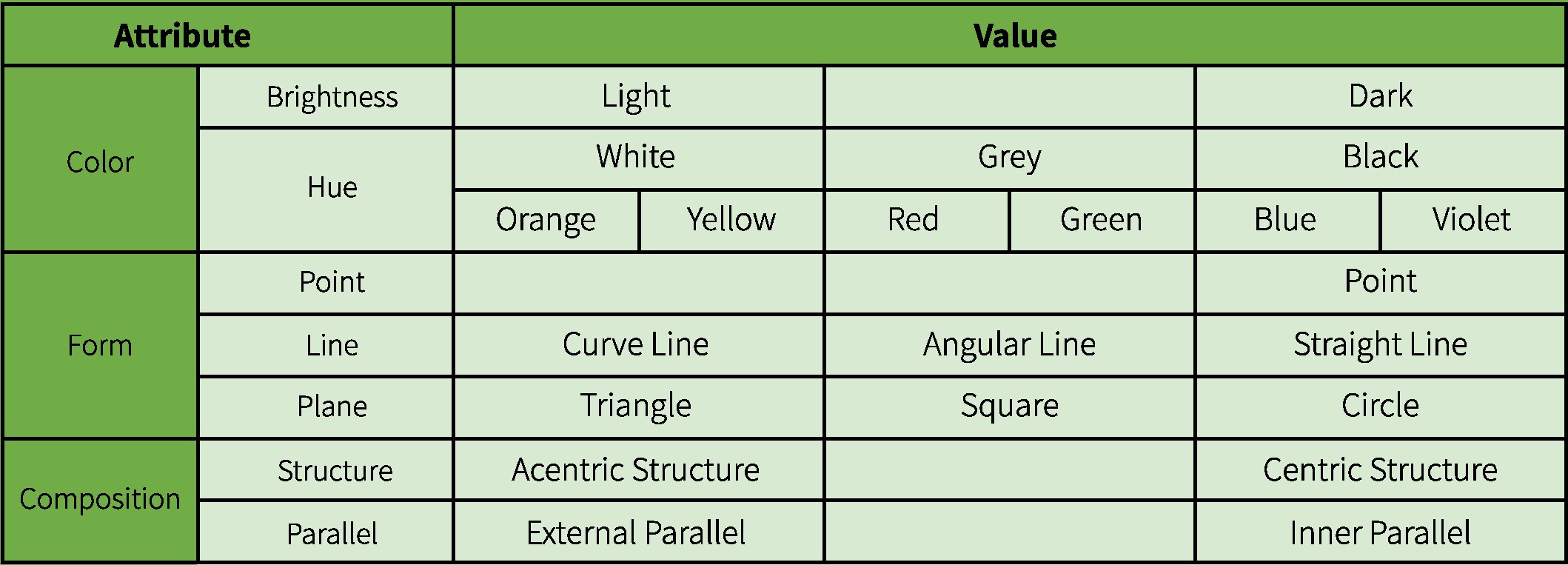}
    \caption{Semantic Descriptive Guideline: Attribute-Value.}
    \label{fig:attribute-value}
    \end{figure}

\subsection{Kandinsky Bauhaus Style Semantic Descriptive Guideline} 

To systematically describe Kandinsky's Bauhaus style, we collaborate with researchers specializing in Kandinsky's art and examine the texts in \textit{Concerning the Spiritual in Art} \cite{Kandinsky_1977} and \textit{Point and Line to Plane} \cite{Kandinsky_Rebay_1979} to establish a semantic descriptive guideline. By summarizing the literature and Kandinsky’s theories, we derived a list of attributes and values encompassing Kandinsky’s style. Appendix \ref*{cfc} provides a brief introduction to Kandinsky’s theories on color, form, and composition. Fig \ref*{fig:attribute-value} shows our summarization of the attribute-value list, which consists of seven attributes and twenty-two values.

\subsection{Kandinsky Bauhaus Style Paintings}
    
We gather 209 of Kandinsky's paintings from his Bauhaus period and meticulously narrowed down the selection to 65 representative artworks through expert evaluation and curation. The selection criteria encompass two key aspects: (1) The artworks need to be complete pieces rather than incomplete sketches, and (2) they are required to exhibit explicit attributes that align with the descriptions presented in \textit{Point and Line to Plane} and our semantic descriptive guideline. 

\begin{figure}[h]
    \centering
    \includegraphics[width=0.6\textwidth]{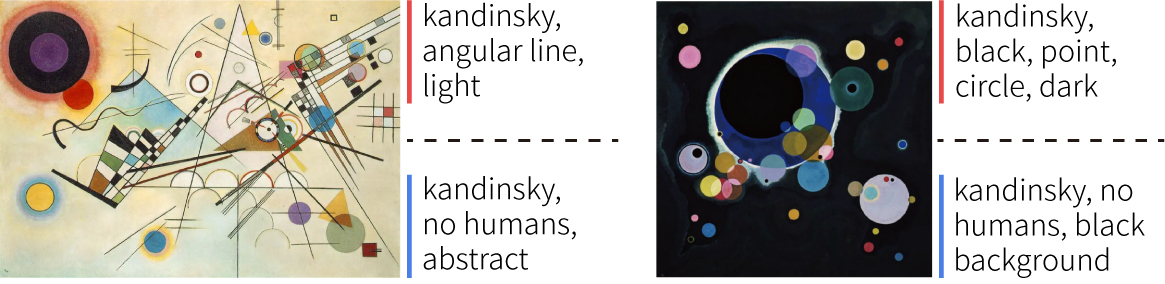}
    \caption{Kandinsky Text-to-Image Dataset.}
    \label{fig:dataset}
    \end{figure}

We collaborate with experts to categorize and label the selected paintings using the attribute-value list. Fig \ref{fig:dataset} shows two examples of text-image pairs, with the red-marked text descriptions derived from our labeling process. The text descriptions start with the prefix ``Kandinsky'' and are followed by the attribute-values. We selectively retain attribute-value with discernible effects while omitting those deemed ambiguous or controversial. This empirical approach ensures a well-balanced dataset across various attributes of Kandinsky Bauhaus features.

\begin{figure}[t!]
\centering
\begin{subfigure}[t]{\textwidth}
    \centering
    \includegraphics[width=\textwidth]{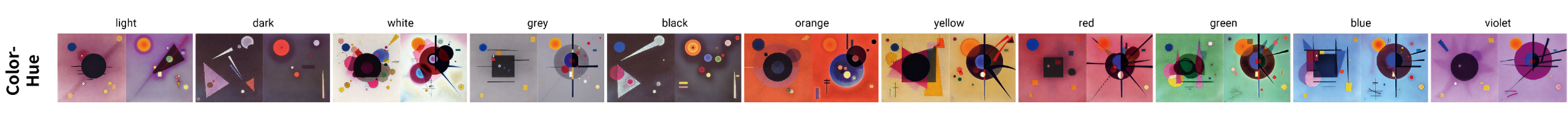}
    \caption{Baseline Model Performance on Hue Attribute.}
    \label{fig:hue_result}
    \end{subfigure}

\begin{subfigure}[t]{\textwidth}
    \centering
    \includegraphics[width=\textwidth]{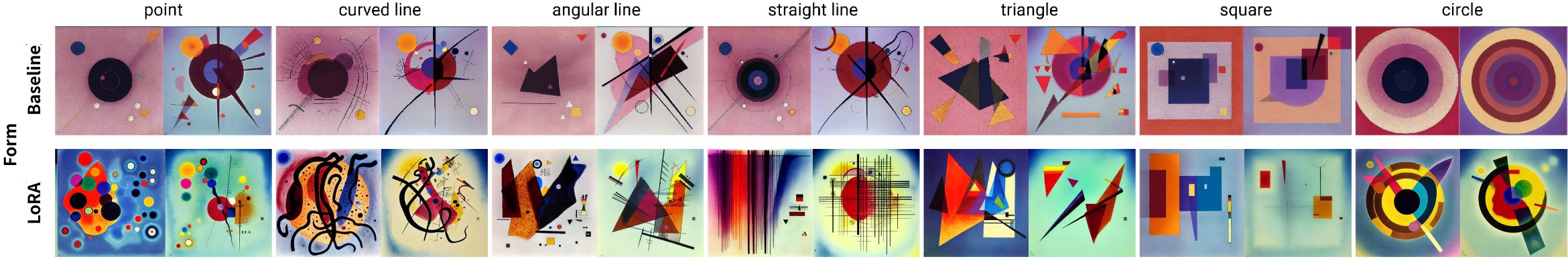}
    \caption{FastLoRA vs. Baseline on Form-Related Attributes.}
    \label{fig:fastlora_result}
    \end{subfigure}

\begin{subfigure}[t]{\textwidth}
    \centering
    \includegraphics[width=\textwidth]{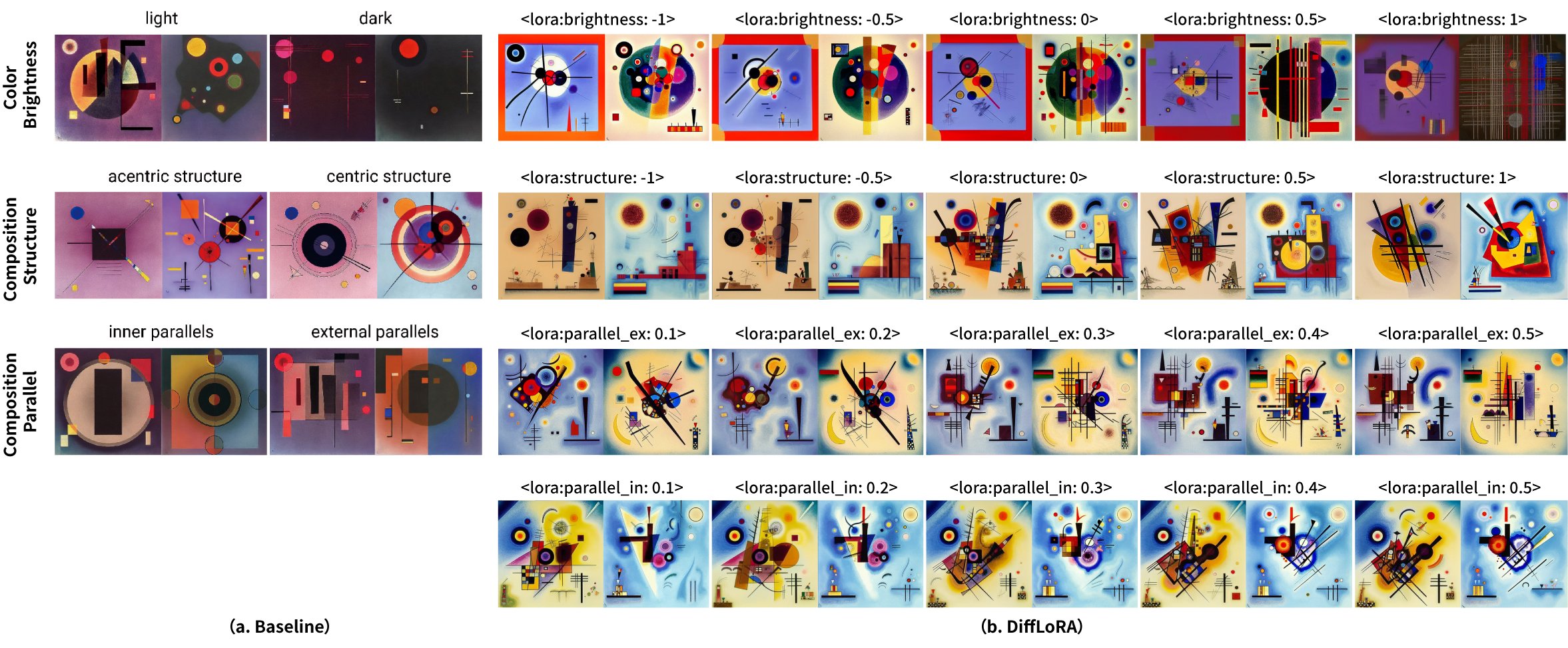}
    \caption{DiffLoRA vs. Baseline on Brightness and Composition-Related Attributes.}
    \label{fig:difflora_result}
    \end{subfigure}
\caption{Comparison of Baseline and the Artist Models from Semantic Injection.}
\label{fig:single-word-prompting}
\end{figure}

\section{Artist Model through Semantic Injection}\label{artistmodel_exp}

Stable Diffusion is one of the most popular open-source solutions for text-to-image generation due to its powerful features and user-friendly interface \cite{Rombach_Blattmann_Lorenz_Esser_Ommer_2022}. However, like most large models, it is not built for specific artists or styles, such as Kandinsky Bauhaus style. To address this, we utilize the standard Stable Diffusion model \cite{Runwayml} and employ semantic injection for customizing the style. 

To semantically inject discrete values like form-related attributes (e.g., point, line, and plane) and hue, we employ the FastLoRA technique. Although hue is generally considered a continuous value, Kandinsky categorizes it into six primary colors, necessitating us to treat it as a discrete value. For continuous attributes, such as brightness and composition-related features, we use DiffLoRA to inject semantic information and derive three additional LoRA models accordingly. By combining these four LoRAs, we complete the semantic injection of Kandinsky's attribute-value list. 

Additionally, we construct a baseline model to facilitate a comparative evaluation of the proposed method. We initially employ an automatic labeling extension \cite{Toriato} from the TTI-art community to annotate the collected Kandinsky paintings. The baseline model was fine-tuned with these automatically labeled descriptions as shown in Fig \ref*{fig:dataset}. Subsequently, we apply the semantic injection method to fine-tune the Stable Diffusion model using both the automatically labeled data and our Kandinsky Bauhaus style dataset, resulting in two models: the baseline and our proposed Artist Model.
    
\subsection{Hue and Form-related Attributes}

To evaluate the performance of our FastLoRA models, we use single-word prompts derived from the attribute-value list and compare the generated results with the baseline model. Fig \ref{fig:hue_result} demonstrates the effective performance of the baseline model on the color hue attribute, leading us to regard the color hue attribute as inherent to the original model. Due to space limitations, we choose to omit the similar results produced by the fine-tuned fast LoRA model. Fig \ref*{fig:fastlora_result} highlights a comparative analysis of the baseline model and our FastLoRA model on form-related attributes, demonstrating the significantly superior performance of FastLoRA in generating Kandinsky-specific form features.

\subsection{Brightness and Composition-related Attributes.}
    
\begin{figure}[h]
    \centering
    \includegraphics[width=\textwidth]{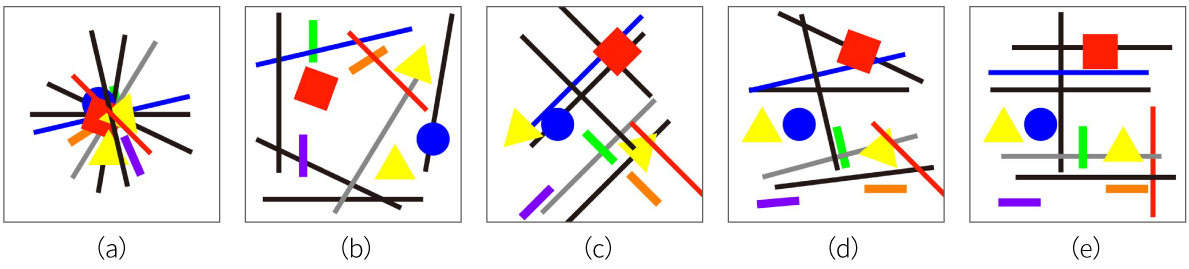}
    \caption{Data Sample for Composition-related Attributes.}
    \label{fig:difflora_data}
    \end{figure}

To incorporate the brightness attribute, we select a Kandinsky painting as the reference image (Image A), adjust its brightness to create a much darker version (Image B), and then obtain the DiffLoRA for brightness. For the structure and parallel composition attributes, modifying an existing painting to reflect these characteristics is challenging. To overcome this, we manually create the inputs representing these attributes as in Fig \ref*{fig:difflora_data}. Kandinsky's definition of inner parallel composition as diagonal and external parallel composition as edges deviates from conventional understanding, making it difficult for the model to learn these attributes. To address this, we introduce Fig \ref*{fig:difflora_data}d as an intermediate input representing parallel structures and pair it with two extreme cases-Fig \ref*{fig:difflora_data}c and Fig \ref*{fig:difflora_data}e-representing inner and external parallel composition, respectively. This approach results in creating two DiffLoRA models that effectively capture the attributes. Finally, Fig \ref*{fig:difflora_result} demonstrates that our DiffLoRA model successfully synthesizes brightness and composition-related features using single-word prompts. 

\subsection{The Artist Model}

We combine the fine-tuned FastLoRA and DiffLoRA models to develop the Artist Model. To validate its generative capabilities, we collaborate with a Kandinsky expert. Using prompts aligned with our semantic descriptive guideline, the expert prompts the Artist Model to generate Kandinsky Bauhaus-style abstract art. Fig \ref{fig:prompting}a illustrates a prompt and its corresponding generated image, which was based on the expert's vision of a painting featuring ''warm-temperature colors (e.g., orange), light tones (e.g., white, grey), and neutral form elements (such as points, angular lines, and squares) parallel to the edges of the canvas in an acentric structure.'' Fig \ref{fig:prompting}b and Fig \ref{fig:prompting}c showcase two additional examples. For users familiar with our semantic descriptive guideline or Kandinsky’s theories, the Artist Model allows them to generate Kandinsky Bauhaus-style tailored to their preferences. However, this controllable process still relies on prompting and has not yet achieved our envisioned goal of prompt-free generation.

\section{Genetic Prompting Optimization and Visualization}\label{promptingmodel_exp}

\subsection{Procedural Prompt Generation}

Using our semantic descriptive guideline for the Kandinsky Bauhaus style, we implement a procedural prompt generation function associated with the Artist Model. The prompting of the artist model encompasses attribute-value pairs and the names of the DiffLoRA models. For attributes with discrete values, such as hue and form-related attributes, we use FastLoRA models, with these attributes collectively represented as $A_M$, which includes multiple discrete values. When prompting attributes with continuous values, such as brightness and composition-related attributes, the attributes are denoted as $A_Q$. Due to the distinct types of these attribute-values, we employ Heterogeneous Encoding to represent the information in prompts as genes. The chromosome is described as $C={\left\{Style,A_{Q},A_{M},S \right\}}$. $C$ represents gene positions for different attributes. The term $Style$ indicates the consistent keyword, e.g., Kandinsky, in the prompt, while $S$ represents the random seed of Stable Diffusion. In this experiment, we initialize a set of seeds at the first generation by assigning integers within the range $[0,2147483647)$.

\subsection{Genetic Optimization}

This process begins by forming an initial population of $n$ individuals. The user votes for their preferred individuals and the fitness function evaluates the fitness of each individual to select survivors for the next generation. The genetic mechanism produces a new population of individuals using the selected survivors. This iterative process continues until the user is satisfied with the generated images, resulting in an Optimized Prompting Model.

\subsubsection{Selection} 

We use real-time human feedback (i.e., the current user's votes) to select individuals for the next generation based on their fitness value.

\begin{itemize}
    \item \textbf{Fitness Function:} 
    In each iteration, the user votes for individuals within a population of $n=16$. The number of votes received by an individual is denoted as $V_i$, where $i\in(1,16)$, and $V_i$ is an integer in the range of $(0,+\infty)$. The fitness function is defined as $f(i)=V_i$. We assign weights to the discrete values of attributes $A_M$ for optimization.
    
    \item \textbf{Weight Updating:}
    Initially, all possible values have weights set to 1, denoted as $w_{v}=1$, where $v$ represents one attribute-value in the prompt. In each iteration, the updated weight is computed as $w_{v}'=w_{v}+\sum^{n}_{i=0}V_i$, where $v \in C_{i}$,  $C$ is the chromosome (prompt) for individual $i$. Assuming a consistent user preference, attributes with continuous values are modeled using a normal distribution. User feedback is applied to update the mean and variance of the normal distribution, reflecting the dominant characteristics of the population over iterations.
\end{itemize}

\subsubsection{Crossover} 
For crossover, we select two parents using the Roulette Wheel Selection \cite{Holland_1992}. The selection probability for an individual is calculated as $P_i =\frac{F_i}{\sum^{n}_{j=1}F_j}$, where $n$ represents the population size. We use different strategies to combine genetic information from two parents into offspring.

\begin{itemize}
    \item For $Seed$ gene, we employ Uniform Crossover \cite{Leong_2017} with a fixed probability of $p=0.5$ to determine which parent's gene is inherited. 

    \item For $A_M$, a without-replacement drawing from possible values is performed. 

    \item For $A_Q$, the offspring's value is generated using Average Crossover based on the values of both parents.
\end{itemize}

We generate only one offspring per crossover due to the following reasons:
\begin{itemize}
    \item \textbf{User Interaction:} It ensures a moderate population size, preventing user fatigue during selection.
    \item  \textbf{Methodological Suitability:} Since Uniform and Average Crossover generate genes independently, producing a single offspring does not reduce their effectiveness.
    \item \textbf{Population Diversity:} With a small population size (n = 16), creating one offspring per crossover increases diversity by allowing more unique parent combinations.
\end{itemize}

\subsubsection{Mutation} 
We employ the Uniform Mutation for discrete-valued attributes ($A_M, Seed$) with a mutation rate of $p=0.05$. Each attribute's value is randomly modified with a new value from its possible set. The values of muted $A_M$ are selected without replacement. For the gene encoded as real numbers ($A_Q$), new values are sampled from a normal distribution based on the mean and variance of successful traits from previous iterations. This strategy ensures that the evolutionary history guides mutations toward promising solutions.

\begin{figure}
    \centering
    \includegraphics[width=0.8\textwidth]{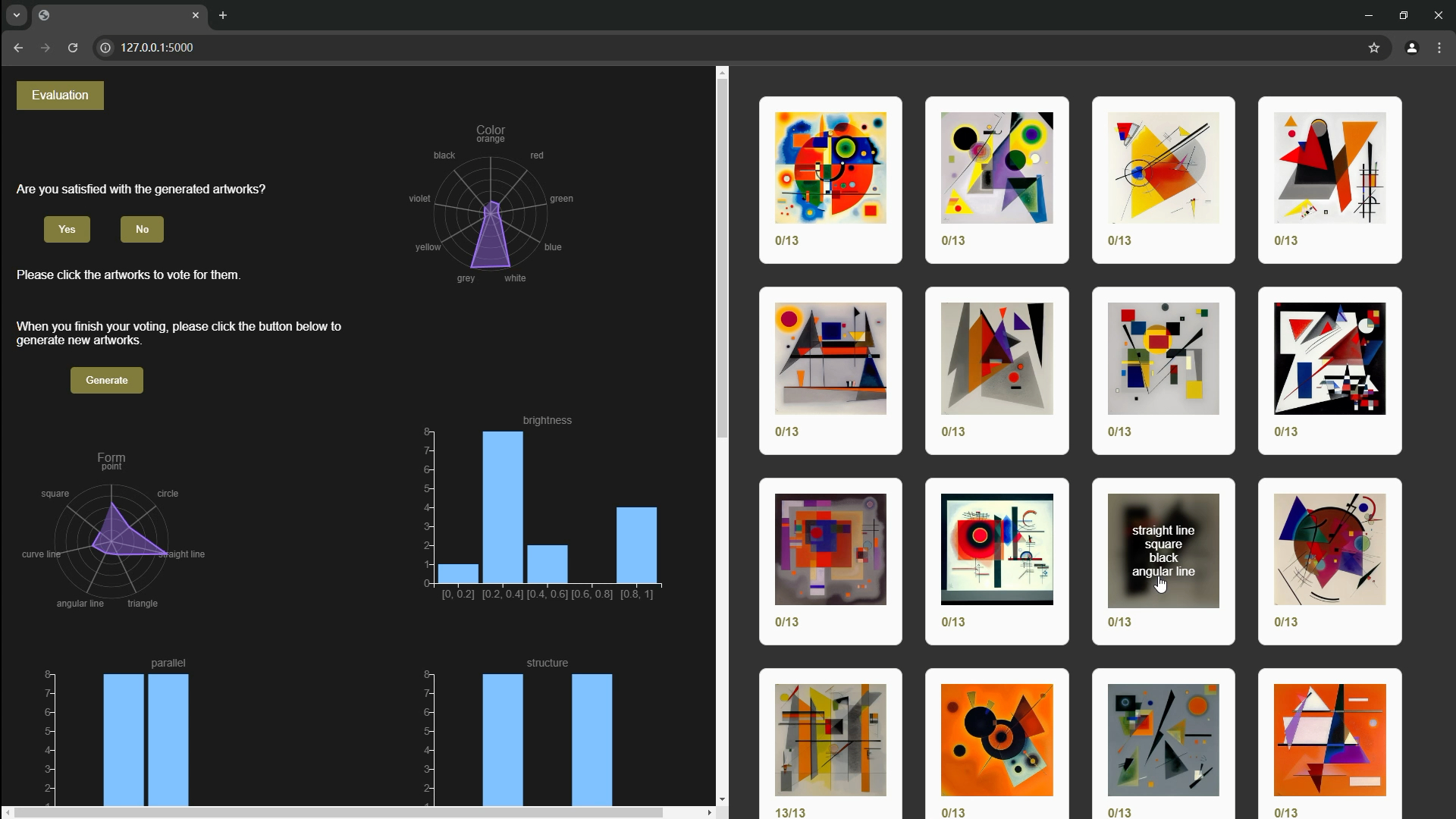}
    \caption{The Interface Design}
    \label{fig:interface}
\end{figure}

\subsection{Prompt Visualization: Iterative Attribute-Value Distribution}
We integrate several visualization techniques, such as radar chart, bar chart, and stream graph, to present the prompt iterations, enhancing users' understanding of their preferences in both image and text domains. Radar charts are used to represent attributes with discrete values, such as Hue and Plane, while bar charts depict attributes with continuous values, showcasing the distribution of attribute values in the current user-voted and selected prompts. Stream graphs illustrate the progression of attribute-value distributions over multiple iterations, providing users with a clear view of evolving preferences. These visualizations facilitate a deep exploration of how attribute-value aligns with individual preferences, making the iterative process transparent, engaging, and comprehensible.

\subsection{The Interactive System and Prompting-free Experience}\label{experience}

\begin{figure}[h]
    \centering
    \includegraphics[width=\textwidth]{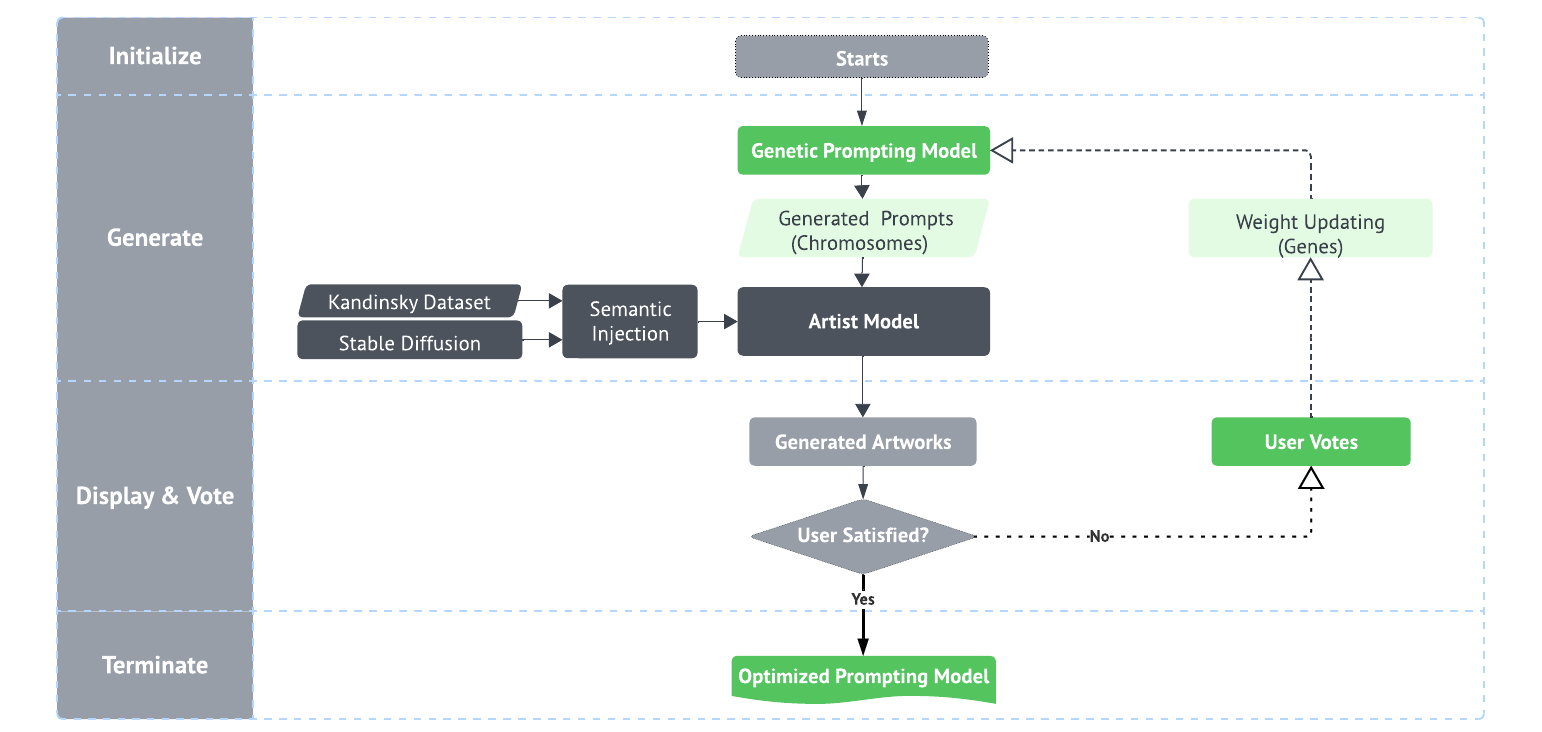}
    \caption{Flowchart: the Interactive System for Prompting-free Experience}
    \label{fig:flowchart}
\end{figure}

Combining the Artist Model and genetic prompting optimization, we implement an interactive system with a point-click interface (as in Fig \ref{fig:interface}) that enables end users to obtain their Optimized Prompting Model in real time and generate their preferred abstract art. Fig \ref{fig:flowchart} illustrates the interaction workflow. The system initializes with a random prompts set based on the original genes pool, and the user views the generated artwork. Optionally, users can hover over a generated image to display its associated prompt and examine the attribute-value distributions through the visualization graphs displayed on the side.

If the user finds the generated artwork misaligned with their preference, they can iteratively vote for their preferred abstract art images by clicking on the images. The votes guide the optimization of the prompt generation until the system produces an Optimized Prompting Model that satisfies the user. Fig \ref{fig:interface} demonstrates the interface design, including the generated artworks and the visualizations of attribute-value distributions throughout the iterative process.

During testing within the research team, convergence is typically achieved within 3 to 5 iterations, often taking less than 5 minutes. This user-friendly process allows users to almost effortlessly obtain an Optimized Prompting Model reflecting their preferences, enabling automatic generation of their desired Kandinsky Bauhaus-style abstract art without the need for explicit prompting.

\section{Discussion and Conclusion} \label{discussion}

\subsubsection{Art Experiences as Solution}
This paper addresses the challenges faced by the artist community when working with large text-to-image models. We propose and implement an art-centered approach that draws heavily on the collective experiences and established techniques from the TTI-art community, as well as classic methods from Generative Art, such as procedural modeling and genetic algorithms. These artistic experiences empower us to exert creative control over large text-to-image models for abstract art synthesis. By exploring the evolution of Generative Art and its potential impact on technology, we provide valuable insights into how art practices can improve collaboration with large AI models.

\subsubsection{Mixing Rule-driven and Preference-based Strategies}
Our two-part approach begins with semantically injecting Kandinsky’s theory into the large model and then applies a genetic algorithm to incorporate the user’s personal preferences. This process balances expert knowledge with user personality by combining rule-driven and preference-based mechanisms. The interplay between these approaches strikes a delicate balance between total control and creative unpredictability, achieving optimal results that blend high visual quality with an emotional connection to the user’s preferences.

\subsubsection{Steering AI as New Artist Strategy}
As AI models continue to grow in complexity and size, artists are increasingly unable to train or recreate models due to resource constraints. Additionally, the environmental costs, such as the carbon footprint of training large models, present significant challenges to building models from scratch. Our proposed approach exemplifies a paradigm shift for artists working with large AI models, advocating for ``steering AI'' as a new strategy that emphasizes fine-tuning, tweaking, and building upon existing models to offer an expressive and eco-friendly solution for art practices with AI.

\section{Limitation and Future Work} \label{limitation}
The presented work has limitations that should be acknowledged. The current experiments are primarily artist and expert-oriented, and focus heavily on technical development with limited user evaluation. To further validate the concept, we plan to conduct a comprehensive user study involving participants with minimal knowledge of Kandinsky's art. This study will aim to collect and analyze users' perceptions and insights regarding the approach. Additionally, the proposed method relies significantly on expert knowledge and manual literature review to establish the semantic descriptive guideline. Future research could focus on automating this process using natural language processing (NLP) techniques to improve efficiency and scalability.



%
%
%
\bibliographystyle{splncs04}
\bibliography{ijcai24}
%






\appendix

\section{Color, Form and Composition Theory} \label{cfc}
\subsubsection{Color}
Kandinsky's color theory consists of three primary colors (i.e., \textit{Red}, \textit{Yellow}, and \textit{Blue}) and three primary tones (i.e., \textit{Black}, \textit{White}, and \textit{Grey}).  The intersection of these primary colors produces \textit{Orange}, \textit{Green}, and \textit{Violet}. He categorizes these six colors into warm and cold temperature, with red, yellow, and orange considered warm, and green, blue, and purple considered cold.  He defines the tone of a painting as light and dark; determined by the proportion of black and white in a composition. Each color represents a unique spiritual emotion and has an influence on others.\cite[pp.39-43]{Kandinsky_1977}

\subsubsection{Form}
Kandinsky presents a unique theory of tension that describes an element based on its shape, position and orientation. He begins with the concept of the \textit{Point} and derives two primary forms: Line and Plane \cite[pp.23-54]{Kandinsky_Rebay_1979}. Lines are influenced by different types of tension and can take the form of \textit{Straight Lines}, \textit{Curve Lines}, or \textit{Angular Lines} \cite[pp.55-112]{Kandinsky_Rebay_1979}. Similarly, planes are influenced by tension and can manifest as \textit{Triangles}, \textit{Squares}, or \textit{Circles} \cite[p.74]{Kandinsky_Rebay_1979}. These seven elements constitute the primary forms in Kandinsky's paintings.

\subsubsection{Composition}
Kandinsky claims interrelationships of all elements influence the composition. The composition involves the positioning of individual elements, the connections between them, and the arrangement of element clusters. Our study aims to extract easy-to-use compositions to guide the model generation and simplify Kandinsky's complex definitions into two compositional relationships: \textit{Acentric} and \textit{Centric Structure} \cite[pp.137-139]{Kandinsky_Rebay_1979}. These relationships describe the tension of elements relative to the center. In contrast, the tension of elements in relation to the composition's boundaries manifests as \textit{Inner} and \textit{External Parallel}, which is parallel to diagonal and edge of the painting respectively \cite[pp.130-131]{Kandinsky_Rebay_1979}.

\end{document}